\documentclass[10pt,letterpaper]{article}
\usepackage{opex3}
\usepackage{amssymb}
\begin{document}
%%%%%%%%%%%%%%%%%% title page information %%%%%%%%%%%%%%%%%%
\title{Localized photonic bandedge modes and orbital angular momenta of light in a golden-angle spiral}

\author{Seng Fatt Liew,$^{\bf{1}}$ Heeso Noh,$^{\bf{1}}$ Jacob Trevino,$^{\bf{2}}$ Luca Dal Negro,$^{\bf{2,3},\dagger}$ Hui Cao$^{\bf{1,4},*}$}

\address{$^1$Department of Applied Physics, Yale University, New Haven, CT 06511, USA\\
$^2$Division of Materials Science and Engineering, Boston University, Brookline,  MA 02446, USA\\
$^3$Department of Electrical and Computer Engineering \& Photonics Center, Boston University, 8 Saint Mary's Street, Boston, MA, 02215, USA\\
$^4$Department of Physics, Yale University, New Haven, CT 06511, USA\\
}

\email{*hui.cao@yale.edu,$^\dagger$dalnegro@bu.edu} %% email address is required

% \homepage{http:...} %% author's URL, if desired

%%%%%%%%%%%%%%%%%%% abstract and OCIS codes %%%%%%%%%%%%%%%%
%% [use \begin{abstract*}...\end{abstract*} if exempt from copyright]

\begin{abstract}
We present a numerical study on photonic bandgap and bandedge modes in the golden-angle spiral array of air cylinders in dielectric media.  Despite the lack of long-range translational and rotational order, there is a large PBG for the TE polarized light. 
Due to spatial inhomogeneity in the air hole spacing, the bandedge modes are spatially localized by Bragg scattering from the parastichies in the spiral structure.  They have discrete angular momenta that originate from different families of the parastichies whose numbers correspond to the Fibonacci numbers.  The unique structural characteristics of the golden-angle spiral lead to distinctive features of the bandedge modes that are absent in both photonic crystals and quasicrystals. 
\end{abstract}

 % REPLACE WITH CORRECT OCIS CODES FOR YOUR ARTICLE

%%%%%%%%%%%%%%%%%%%%%%% References %%%%%%%%%%%%%%%%%%%%%%%%%

%%%%%%%%%%%%%%%%%%%%%%%%%%  body  %%%%%%%%%%%%%%%%%%%%%%%%%%
\section{Introduction}

Golden-angle spirals have been discovered in the arrangements of seeds, leaves, and stalks in sunflowers, pine cones, artichokes, celery, daisies, and many other plants \cite{stevens}.  Such patterns give the most even distributions of seeds in the sunflower heads, with no seeds clumping.  Mathematically the golden-angle spiral is a form of Fermat's spiral representing the densest packing of identical circles within a circular region.  Those circles form many spiral arms, or parastichies, in clockwise (CW) and counter-clockwise (CCW) directions.  The numbers of parastichies are consecutive numbers in the Fibonacci series, the ratio of which approximates the golden ratio \cite{naylor}.  
Inspired by nature, optical properties of spiral structures have been explored in recent years.  
For instances, photonic crystal fibers (PCF) with air holes arranged in the golden-angle spiral pattern exhibit large birefringence with tunable dispersion~\cite{agrawal}.  
Nanoplasmonic spirals generate polarization-insensitive light diffraction and planar scattering over a broad frequency range \cite{luca}. \\

\indent Another fascinating feature of the golden-angle spiral structure is its ability to create an isotropic  photonic bandgap (PBG), which inhibits light propagation in all directions \cite{pollard}.  
The most well-known structures that produce PBGs are photonic crystals \cite{joannopoulos}, but their structural anisotropy leads to spectral mismatch of gaps in different directions.  
To have complete PBGs, more isotropic structures, e.g.  photonic quasicrystals with higher rotational symmetries, are preferred ~\cite{chan,steinhardt}.  
However, the photonic quasicrystals still have discrete Fourier spectra and are not fully isotropic.  
The golden-angle spiral has better isotropy because its Fourier space is diffuse and circularly symmetric \cite{luca}.  
It has been predicted \cite{pollard} that a 2D golden-angle spiral array of dielectric cylinders in air, even with low refractive index contrast, can create a broad omnidirectional PBG for transverse magnetic (TM) polarization.  
The gap width exceeds that in a six-fold lattice or a 12-fold fractal tiling. 
One advantage over the photonic amorphous structure which can also produce an isotropic PBG is that the golden-angle spiral structure is deterministic and has predictable and reproducible properties.  
The absence of sample to sample variations is critical to many applications. \\

\indent Although it is now known that the golden-angle spiral can produce an omnidirectional PBG, little is known on the nature of its photonic bandedge modes.  
In photonic crystals, the photonic bandedge modes have low group velocities and high quality factors, thus useful to slow light devices and lasers.  
The bandedge modes are spatially extended in the photonic crystals, but can be critically localized in the photonic quasicrystals which lacks translational symmetry ~\cite{chan,steinhardt,luca2}.  
The golden-angle spiral does not have discrete translational or rotational symmetries, and its bandedge modes are distinct from those in photonic crystals and quasicrystals.  
Recent studies demonstrated that spiral structures can transfer net orbital angular momentum to the scattered optical waves \cite{luca}.  
Hence, the unique structural characteristic of the golden-angle spiral may impose unique and novel features on the photonic bandedge modes.  \\

\indent In this paper, we present a systematic study on the photonic bandedge modes in 2D golden-angle spiral arrays of air holes in dielectric host media.  
The PBG exists for the transverse electric (TE) polarization, and multiple classes of bandedge modes are identified.  
Each class is localized in a specific region of the structure, due to spatial inhomogeneity in the distribution of neighboring air holes.  
We discover that the photonic bandedge modes possess discrete angular momenta that correspond to the Fibonacci numbers, which are associated with the parastichies in the spiral structure.  
The close relationship between the structural properties and characteristics of the photonic bandedge modes is unveiled using the Fourier Bessel spatial analysis.  
The unique properties of the photonic bandedge modes in the golden-angle spiral may lead to applications in light emitting devices and optical sensors.\\

\section{Structural analysis of the golden-angle spiral}

The golden-angle spiral, also called the Vogel's spiral, was first proposed by Vogel to simulate the seeds distribution in a sunflower head \cite{vogel}.  
The location of each seed or circle is specified by a simple generation rule and expressed in the polar coordinate $(r,\theta)$ as
\begin{eqnarray}
r &=& b \sqrt{q} \, , \label{generation1}\\
\theta &=& q \alpha \, ,\label{generation2}
\end{eqnarray}
where $q = 0, 1, 2, \dots$ is an integer, $b$ is a constant scaling factor, $\alpha = 360^\circ/\phi^2 \approx 137.508^\circ$ is an irrational number known as the ``golden angle'', $\phi = [1+5^{1/2}]/2 = 1.6180339\dots$ is the golden ratio.  
The value of $\phi$ is approached by the ratio of two consecutive numbers in the Fibonacci series $(1, 2, 3, 5, 8, 13, 21, 34, 55, 89, 144, \dots)$.  
With this generation rule, the $q$th circle is rotated azimuthally by the angle $\alpha$ from the location of the $(q-1)$th one, and also pushed radially away from the origin by a distance $\Delta r = b \left( \sqrt{q} - \sqrt{q-1} \right)$.\\

\begin{figure}[htbp]
\centering
\includegraphics[scale = 0.5]{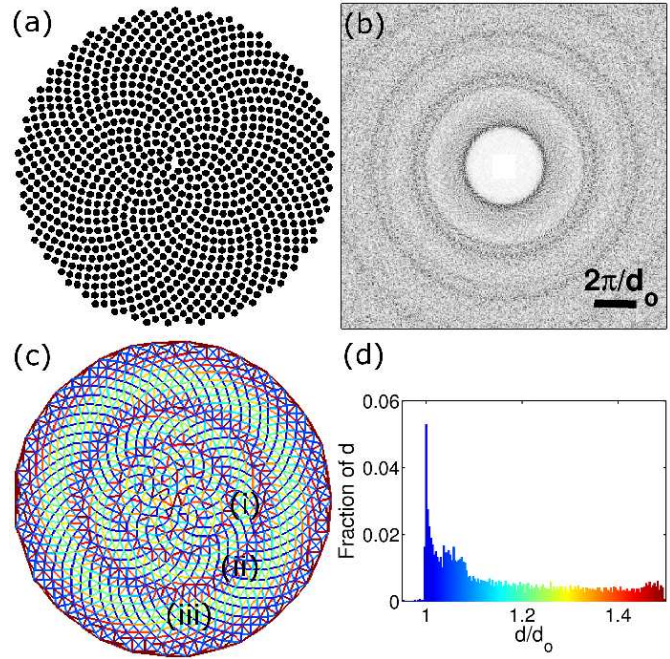}
\caption{(a) Golden-angle spiral array consisting of 1000 circles. (b) Spatial Fourier spectrum of the spiral structure in (a). (c) Delaunay triangulation of (a). The line segments that connect neighboring circles are color-coded by their lengths $d$. (d) Statistical distribution of the distance between neighboring particles $d$ normalized to the most probable value $d_o$. The colors are consistent to those in (c). }
\label{fig1}
\end{figure}

\indent Figure \ref{fig1}(a) shows a golden-angle spiral that consists of $N = 1000$ circles. 
Visually there are multiple families of spiral arms formed by the circles.  
Within each family, the spiral arms, also called parastichies, are regularly spaced.  
Some of the families have parastichies all twisting in the CW direction, and the others in the CCW direction.  
The families are all intertwined.  
The number of parastichies in every family is a Fibonacci number ~\cite{naylor}. \\

\indent The 2D spatial Fourier spectrum of the golden-angle spiral is shown in Fig. \ref{fig1}(b).  
It has a continuous background, on top of which are discrete concentric rings.  
The isotropy of the Fourier space reflects the structural isotropy.  
The diffuse background in the Fourier space indicates the golden-angle spiral is not a quasicrystal, but has much more spatial frequency components \cite{luca}.  
The radii of discrete rings correspond to the dominant spatial frequencies of the structure.  
We extract the distance between neighboring particles $d$ by performing the Delaunay triangulation on the spiral array.  
In Fig. \ref{fig1}(c), each line segment connects two neighboring circles, and its length $d$ is color coded. 
The statistical distribution of $d$ in Fig. \ref{fig1}(d) is broad and non-Gaussian. 
Instead of the average value of $d$, we use the most probable value $d_o$ at which the distribution is peaked to normalize $d$.   
The broad distribution of $d$ is consistent with the rich Fourier spectrum.  
The brightest ring in the Fourier space, which is also the smallest, has a radius close to $2 \pi / d_o$.  
The non-uniform color distribution in Fig. \ref{fig1}(c) reveals the spatial variation of neighboring particles spacing in the spiral structure.  
This special type of spatial inhomogeneity is a distinctive feature of the golden-angle spiral, and it has a significant impact on its optical resonances as will be shown later.\\

\begin{figure}[htbp]
\centering
\includegraphics[scale = 0.45]{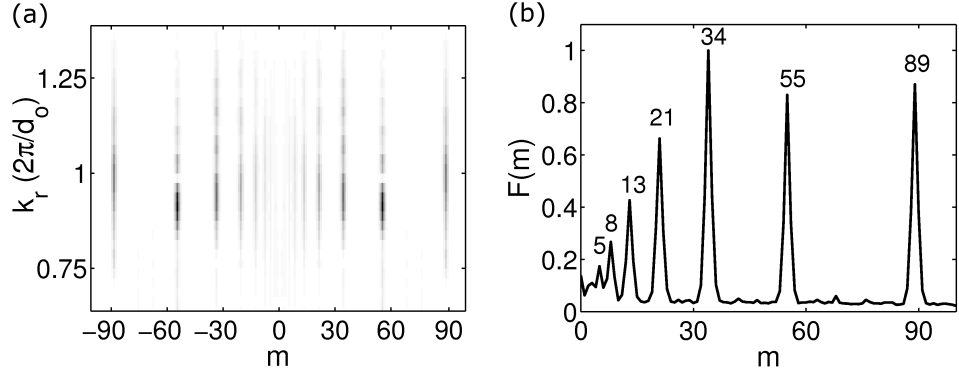}
\caption{Fourier Bessel Transform (FBT) of the golden-angle spiral structure in Fig. 1(a) gives $|f(m,k)|^2$ (a) and $F(m)$ (b). }
\label{fig2}
\end{figure}

\indent In order to better understand the structural complexity of the golden-angle spiral, we resort to the Fourier Bessel spatial analysis.  
The Fourier Bessel transform decomposes the density function associated with the spiral structure in a series of Bessel functions.  
\begin{eqnarray}
f(m,k_r) =  \frac{1}{2\pi}\int_0^{\infty}\int_0^{2\pi}r \, dr \, d\theta \, \rho(r,\theta) \, J_m(k_r r) \,  e^{im\theta} \, , 
\end{eqnarray}
where the density function  $\rho(r,\theta)$ is shown in Fig. \ref{fig1}(a), the azimuthal number $m$ is an integer, and $k_r$ represents a spatial frequency in the radial direction.  
The 2D plot of $|f(m,k_r)|^2$ shown in Fig. \ref{fig2}(a) illustrates that there are multiple and well-defined azimuthal components $m$ in the golden-angle spiral.  
The equal intensity of the positive and negative $m$ components shows that the golden-angle spiral is achiral.  
After integrating over the radial frequency $k_r$, we obtain $F(m)=\int |f(m,k_r)|^2 dk_r$ which is plotted  in Fig. \ref{fig2}(b).  
The frequency range of integration is $[\pi/d_o,3\pi/d_o]$, centered around the dominant spatial frequency $2\pi/d_o$.  
%THE FREQUENCY RANGE of integration is $[\pi/d_o,3\pi/d_o]$, CENTERed AROUND THE DOMINANT SPATIAL FREQUENCY $2\pi/d_o$.
We notice interestingly the dominant $m$ values are 5,8,13,21,34,55,89, which are Fibonacci numbers and represent the number of parastichies in each family.  
Later we will demonstrate that the parastichies encode discrete angular momenta, quantized in the Fibonacci numbers, onto the optical resonances. \\

\section{Photonic bandgap and bandedge modes}

\begin{figure}[htbp]
\centering
\includegraphics[scale = 0.5]{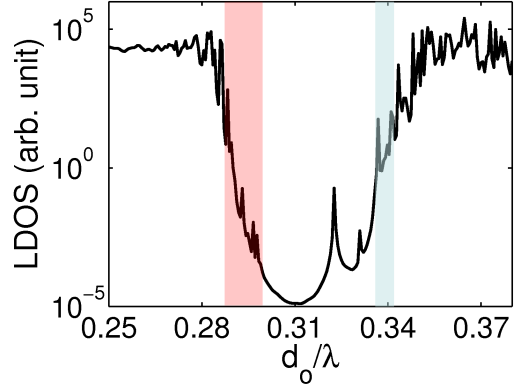}
\caption{LDOS calculated at the center of the golden-angle spiral array as a function of the normalized frequency $d_o / \lambda$. The regions at the lower and upper bandedge where the bandedge modes exist are highlighted. }
\label{fig3}
\end{figure}

We investigate now the optical properties of a golden-angle spiral that consists of $N=1000$ air cylinders in a dielectric medium with refractive index $n=2.65$.  
This structure, inverse of that in Ref. \cite{pollard}, facilitates the formation of PBG for the TE polarized light with $(E_r, E_{\theta}, H_z)$.  
We calculate the local density of optical states (LDOS) at the center of the spiral structure, 
$g(\textbf{r}, \omega) = ({2\omega}/ {\pi c^2}) Im[G(\textbf{r},\textbf{r},\omega)]$, 
where $G(\textbf{r},\textbf{r'},\omega)$ is the Green's function for the propagation of $H_z$ from point $\textbf{r}$ to $\textbf{r'}$.  
The numerical calculation is implemented with a commercial program COMSOL \cite{comsol}.  
From the calculated LDOS in Fig. \ref{fig3}, we clearly see a PBG, and its width is about $11\%$ of the gap center frequency. \\

\begin{figure}[htbp]
\centering
\includegraphics[scale = 0.45]{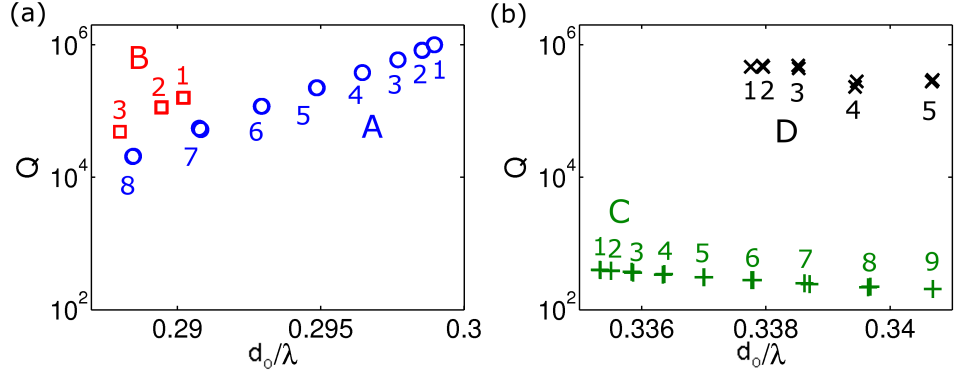}
\caption{Quality factors of the lower bandedge modes (a) and upper bandedge modes (b) versus the normalized frequency $d_o / \lambda$. }
\label{fig4}
\end{figure}

\indent There are two peaks inside the gap at $d_o/ \lambda$ = 0.323 and 0.331.  
They represent defect modes localized at the center of the spiral array where a small dielectric region free of air holes acts as a structural defect.  
At both edges of the gap, there are many more peaks which correspond to the bandedge modes.  
Those on the higher (lower) frequency edge of the gap are denoted as upper (lower) bandedge modes.  
We calculate the complex frequencies $\omega = \omega_r +i\omega_i$ and spatial field distributions of the bandedge modes using the eigensolver of COMSOL.  
The quality factor $Q = \omega_r / 2 \omega_i$ is obtained for every mode.  
From their frequencies and field patterns, we identify several classes of the bandedge modes.  
Within each class the modes have similar field patterns and display monotonic variation of $Q$.  
Two classes of the lower bandedge modes are labeled in a plot of $Q$ vs. $d_o / \lambda = \omega_r d_o / 2 \pi c$ in Fig. \ref{fig4}(a), another two classes of the upper bandedge modes in Fig. \ref{fig4}(b).  
Within each class, the modes are ordered numerically following their spectral distances from the edge of the PBG.  
As the modes in each class move further away from the PBG, the frequency spacing of adjacent modes increases and the $Q$ decreases.\\

\begin{figure}[htbp]
\centering
\includegraphics[scale = 0.1]{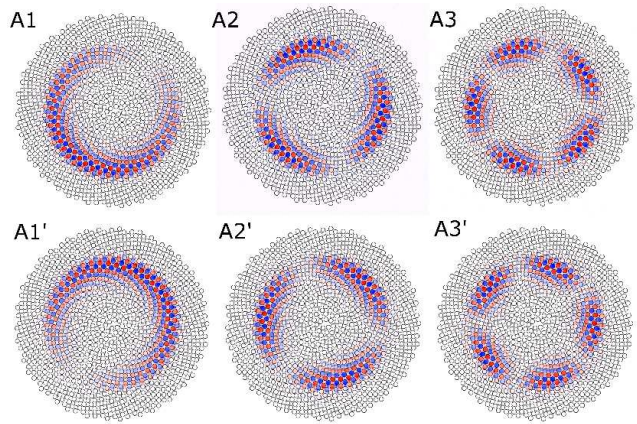}
\caption{ Spatial distributions of magnetic field $H_z$ for the first three pairs of bandedge modes of class A. The modes are localized within a ring of radius $\sim 12d_o$.  }
\label{fig5}
\end{figure}

\begin{figure}[htbp]
\centering
\includegraphics[scale = 0.1]{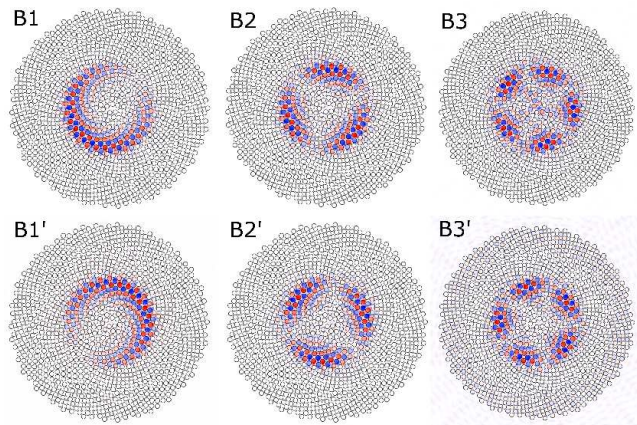}
\caption{ Spatial distributions of magnetic field $H_z$ for the first three pairs of bandedge modes of class B. The modes are localized within a ring of radius $\sim 7d_o$
, close to the center of spiral than the modes of class A. }
\label{fig6}
\end{figure}

\begin{figure}[htbp]
\centering
\includegraphics[scale = 0.11]{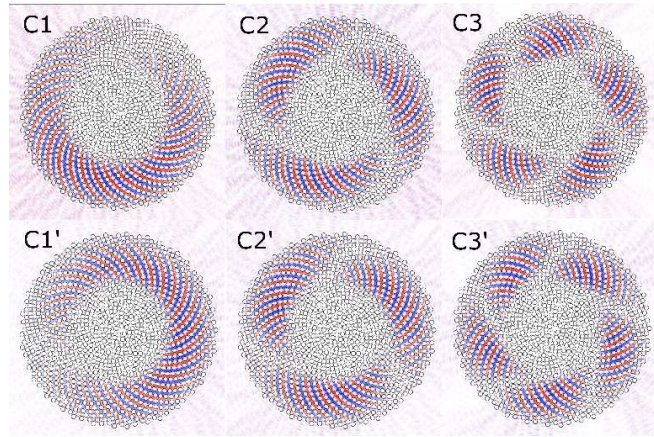}
\caption{ Spatial distributions of magnetic field $H_z$ for the first three pairs of bandedge modes of class C. The modes are located near the boundary of the spiral and have stronger light leakage through the boundary. }
\label{fig7}
\end{figure}

\begin{figure}[htbp]
\centering
\includegraphics[scale = 0.11]{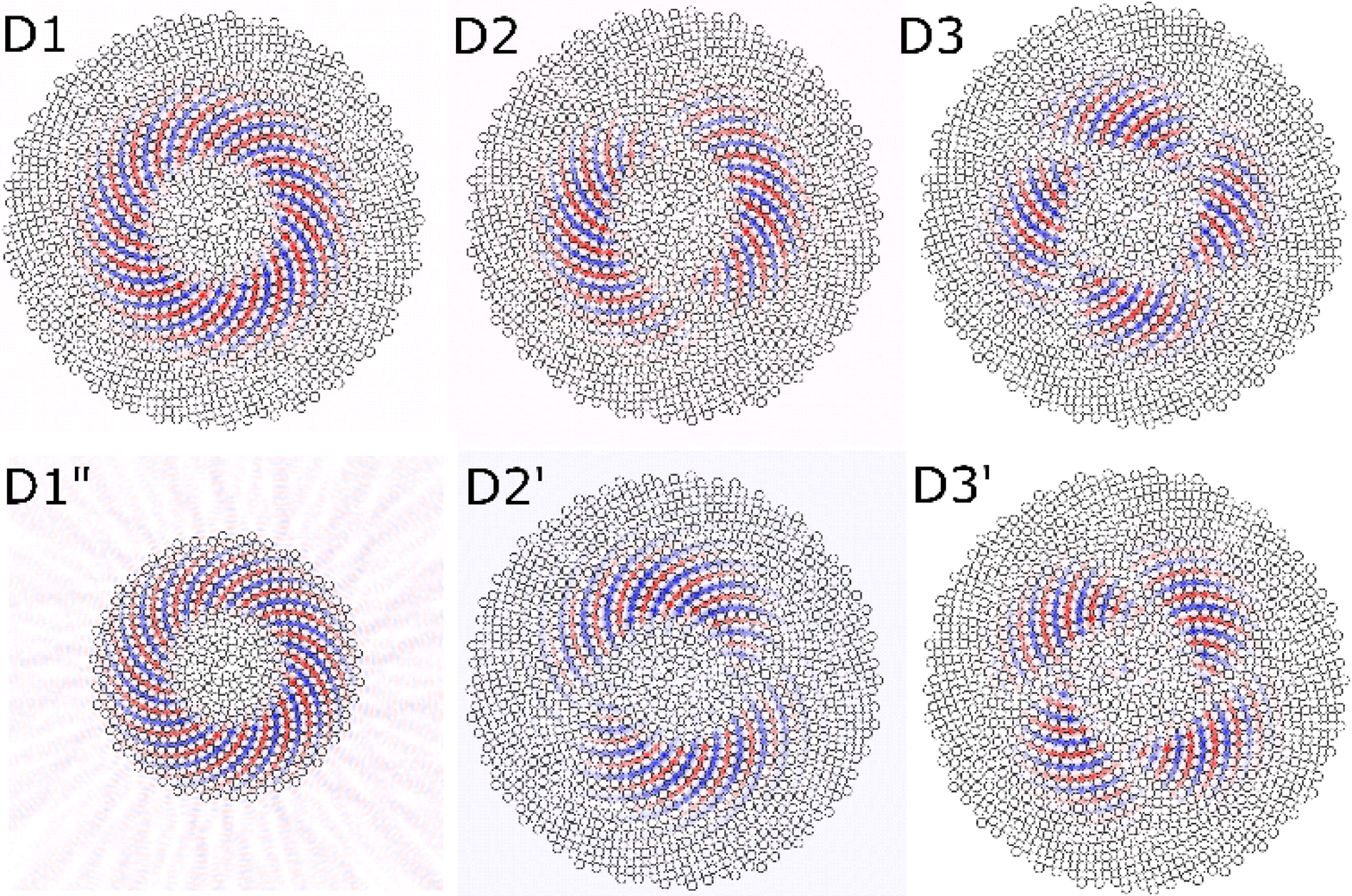}
\caption{ Spatial distributions of magnetic field $H_z$ for the first three pairs of bandedge modes of class D. The modes are localized closer to the center of the spiral and have small light leakage through the outer boundary. Mode D1" has the same field distribution as D1 after the air cylinders in the outer layer of the spiral are removed. The light leakage increases since the mode is closer to the boundary now. The insensitivity of mode D1 to the change at the boundary confirms it is a localized mode.}
\label{fig8}
\end{figure}

\indent The spatial distributions of the magnetic field ($H_z$) for the first three modes in classes A, B, C and D are presented in Figs. \ref{fig5}-\ref{fig8}.  
Every mode is accompanied by a degenerate mode, e.g., A1 and A1' have the same frequency and complementary spatial profile.  
The lower bandedge modes have magnetic (electric) field mostly concentrated in the air (dielectric) part of the structure, while the upper bandedge modes are just the opposite.  
This behavior is similar to that of a photonic crystal, but there are also remarkable differences.  
The bandedge modes in the golden-angle spiral are spatially localized, each class of modes is confined within a ring of different radius.  
As long as the ring is notably smaller than the system size, the modes are insensitive to the boundary, as for the localized states.  
For example, mode D1 remains unchanged when the air cylinders near the boundary are removed [D1" in Fig. \ref{fig8}].\\

\indent A careful inspection of the mode profiles reveals that the class A modes have the magnetic field maxima  along the parastichies that are formed by the air cylinders and twist in the CCW direction, while class B follow a different family of parastichies that twist in the CW direction.  
These local standing wave pattern behaviors indicate light is confined in the direction perpendicular to the parastichies via Bragg scattering from the air holes.  
Since the orientation of parastichies changes with the polar angle, these standing waves rotate azimuthally and wrap around to form a circular pattern.  
Similarly, the magnetic field maxima of class C modes stay along the dielectric parastichies that are formed in between the air cylinders and twist in the CCW direction, and class D on a different family of dielectric parastichies twisting in the CW direction.  
Bragg scattering from the dielectric parastichies leads to the confinement of light in a ring.  
The envelop functions of the bandedge modes exhibit clear modulations in the azimuthal direction.  
The number of nodes (zero points) is an odd integer for the modes in class A, B, C, but an even integer for  D.  
This feature, as will be explained later, is related to the number of parastichies on which the modes are located.\\

\section{Spatial inhomogeneity and localization}

\begin{figure}[htbp]
\centering
\includegraphics[scale = 0.25]{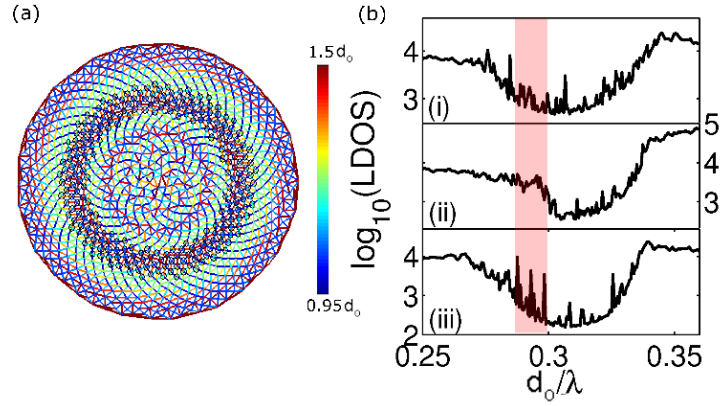}
\caption{ (a) Overlay of the region where class A modes are localized on the color map of the neighboring particles distance of air cylinders revealing class A modes stay mostly inside a ring labeled (ii) and sandwiched between two other rings (i) and (iii).  
(b) LDOS in the regions (i), (ii) and (iii). }
\label{fig9}
\end{figure}

\indent In this section, we will demonstrate that the spatial localization of photonic bandedge modes in the golden-angle spiral structure results from inhomogeneous distribution of spacing between neighboring particles $d$.
From the colors of line segments connecting neighboring circles in Fig. \ref{fig1}(c), we see alternating rings of green color [(i) and (iii) in Fig. \ref{fig1}(c)] and blue-reddish color [(ii) in Fig. \ref{fig1} (c)].  
Different classes of bandedge modes are localized in the rings of distinct colors.  
For example, by overlaying the region that contains $90\%$ energy of modes in class A on the color map of $d$ in Fig. \ref{fig1}(c), we find these modes are confined in region (ii), which is sandwiched by regions (i) and (iii) of different color.  
The distribution of $d$ in region (ii) is distinct from that in (i) or (iii), leading to a change of PBG.  
We compute the LDOS in regions (i), (ii) and (iii) by removing air cylinders outside that region.  
As highlighted in Fig. \ref{fig9}(b), the frequency range of class A modes is inside the PBG of region (i) and (iii) but outside the PBG of region (ii).  
Consequently, light within this frequency range is allowed to propagate in region (ii) but not in (i) or (iii).  
Hence, regions (i) and (iii) act like barriers that confine class A modes in region (ii).\\

\indent Next we consider the upper bandedge modes, e.g. class C modes that concentrate in region (iii).  
The LDOS in region (ii) exhibits little difference from that in (iii) within the frequency range of class C modes.  
Thus region (ii) does not act like a barrier to confine modes in (iii).  
However, in region (iii) the distances between some air cylinders match the wavelengths of class C modes, thus providing distributed feedback for the formation of class C modes.  
Consequently, the class C modes stay mostly in region (ii), even though there is no barrier at the boundary of this region.  
It is similar to the formation of resonances in the conventional distributed feedback structures.  
Thanks to its broad distribution of spacing between neighboring particles $d$, the golden-angle spiral can support numerous modes at different frequencies.  
The spatial inhomogeneity of $d$ leads to mode confinement in different parts of the structure. \\

\section{Discrete angular momentum}

\begin{figure}[htbp]
\centering
\includegraphics[scale = 0.3]{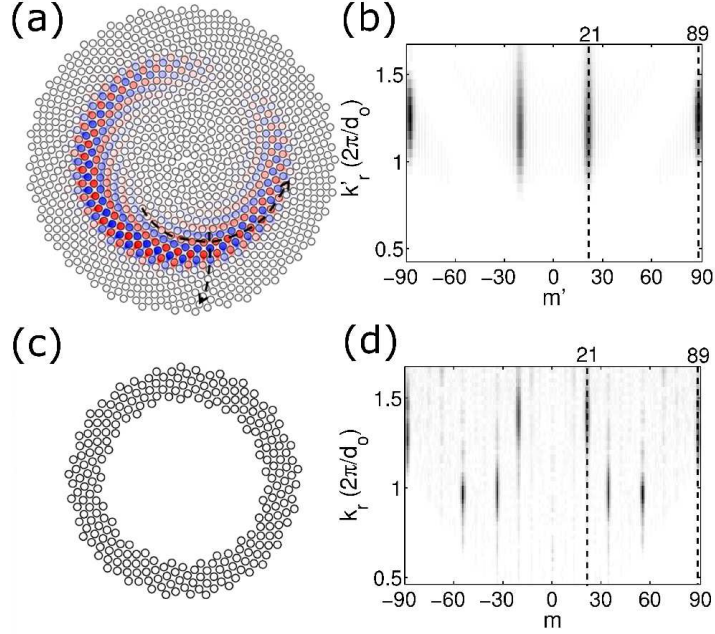}
\caption{ (a) Magnetic field distribution of mode A1 revealing the field maxima follow a family of 21 parastichies twisting in the CCW direction and another family of 89 parastichies in the CW direction (both are marked by the dashed arrows). (b) FBT of the field distribution in (a) gives $F(m',k'_r)$. (c) Region of the spiral array that contains $90 \%$ energy of mode A1 is shown after removing the air cylinders outside. (d) FBT of the structure in (c) gives $F(m,k_r)$ of the local region where mode A1 stays. }
\label{fig10}
\end{figure}

\indent As mentioned earlier, the standing wave patterns of the photonic bandedge modes are formed by distributed feedback from the parastichies that spiral out.  
One example is presented in Fig. \ref{fig10} (a), where the dashed arrows denote two families of parastichies along which the field maxima of mode A1 follow.  
The magnetic field $H_z$ oscillates between the positive maxima on one parastichy and the negative maxima on the next one of the same family.  
We perform the FBT on the field distribution by replacing $\rho(r,\theta)$ in Eq. (3) with $H_z(r,\theta)$.  
To compare with the FBT of the structure [$\rho(r,\theta) > 0$], we set $m'=2m$ and $k'_r=2k_r$ for the field FBT, which is equivalent to considering the FBT of the field intensity distribution. \\    

\indent As shown in Fig. \ref{fig10} (b), mode A1 has discrete angular momenta $m'=21$ and $m'=89$, both are Fibonacci numbers.  
To find their origin, we perform FBT of the structure in the region where mode A1 is localized [Fig. \ref{fig10}(c)].  
The result is presented in Fig. \ref{fig10}(d), and show indeed $m=21$ and $m=89$ components with radial frequency $k_r$ similar to that in the field profile of mode A1.  
While there are also $m=34$ and $m=55$ components in the structure, they are at lower $k_r$, thus corresponding to modes at lower frequencies and further away from the bandedge.  
Hence, these analysis show that the angular momenta of the bandedge modes are imparted by the underlying structure, more specifically, the parastichies in the golden-angle spiral.  
Similar analysis of mode B1 reveals that it supports angular momenta $m=13$ and $m=55$.  
They are also Fibonacci numbers, but smaller than those of mode A1, because mode B1 is localized in a smaller ring that has less number of parastichies. \\

\begin{figure}[htbp]
\centering
\includegraphics[scale = 0.3]{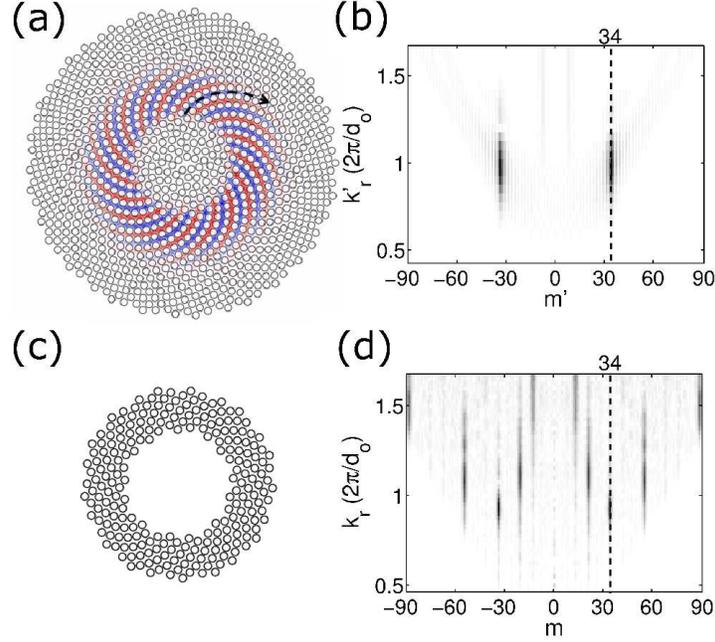}
\caption{ (a) Magnetic field distribution of mode D1 revealing the field maxima follow a family of 34 parastichies twisting in the CW direction(marked by the dashed arrow). 
(b) FBT of the field distribution in (a) gives $F(m',k'_r)$. 
(c) Region of the spiral array that contains $90 \%$ energy of mode D1 is shown after removing the air cylinders outside. 
(d) FBT of the structure in (c) gives $F(m,k_r)$ of the local region where mode D1 stays. }
\label{fig11}
\end{figure}

\indent Moving to the upper bandedge, Figure \ref{fig11}(a) shows that mode D1 is located along the parastichies twisting in the CW direction (marked by the dashed arrow).  
FBT of the mode profile gives a single dominant angular momentum component at $m'=34$.  
FBT of the corresponding region where D1 locates also reveals that there is a $m=34$ component at the similar value of $k_r$.  
Other $m$ components in the structure have higher $k_r$, thus corresponding to higher-frequency modes farther away from the bandedge.  
Similar analysis of mode C1 reveals that it has angular momentum $m'=55$, and the underlying structure contains a family of 55 parastichies. \\

\indent With a better understanding of the connections between the mode profiles and the underlying structures, we can now explain why the numbers of nodes in the envelope functions are either odd or even for all modes belonging to one class.  
Note that the number of parastichies that correspond to mode A1, B1 or C1 is an odd number, but that for D1 is an even number.  
A1, B1 or C1 has one node in the envelop function, while D1 has none.  
As mentioned previously, the field maxima alternate between the positive on one parastichy and the negative on the next.  
After wrapping around one turn ($360^{\circ}$) and returning to the original parastichy, the field maxima must coincide with the one at the original parastichy.  
This is possible when there are an even number of parastichies, e.g. for mode D1.  
For mode A1, B1, or C1, the number of parastichies is an odd number, thus the field maxima would change sign after one turn.  
Since the field maxima of different sign cannot coincide spatially, there must be a radial shift, e. g., the positive maxima of the returning field shift away from the negative maxima of the original field along the parastichy, and there is a field node in between them.  
After a second round trip, there must be another nodal point.  
All these nodal points of the field form a node for the envelop function. \\ 

\begin{figure}[htbp]
\centering
\includegraphics[scale = 0.5]{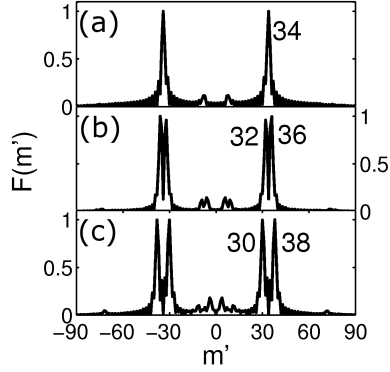}
\caption{ $F(m')$ from the FBT of the field profiles of mode D1 (a), D2 (b), and D3 (c) illustrating the splitting of the peak due to azimuthal modulations of the envelop functions of D2 and D3. }c
\label{fig12}
\end{figure}

\indent  For the higher-order modes in every class of A-D, the number of nodes in the envelop function increases in a step of two, thus remain as an odd number for A-C and an even number for D.  
The addition of an even number of radial shift of field maxima adds a multiple of $2 \pi$ to the phase of the returning field, and does not affect the constructive interference at the starting point.  
We perform FBT on the field distributions of the higher-order modes, and find the additional nodes in the envelope function causes a splitting of the peaks in $F(m')$.  
For example, mode D1 has only a single peak at $m=34$ [Fig. \ref{fig12}(a)], while  
D2 has two peaks at $m=32$ and $m=36$ [Fig. \ref{fig12}(b)].  
The change in $m$, $\Delta m = 2$, is equal to the number of nodes in the envelope function.  
For D3 mode [Fig. \ref{fig12}(c)], $\Delta m = 4$ due to four nodes in the envelop function.  
Such splitting is observed in all higher-order modes of classes A, B and C.  
The azimuthal modulation of the envelop function introduces additional angular momenta to the bandedge modes. \\

\section{Conclusion}

In summary, we have studied numerically the photonic bandgap and bandedge modes in the golden-angle spiral array of air cylinders in dielectric media.  
Despite the absence of long-range translational and rotational order, there exists a significant PBG for the TE polarized light.  
The upper and lower bandedge modes evolve in a deterministic manner, and can be categorized to different classes.  
Due to spatial inhomogeneity in the distances of neighboring air holes, the bandedge modes are localized within the rings of different radii via Bragg scattering from the parastichies in the spiral structure, and wrapped around azimuthally to form circular patterns which carry the well-defined angular momenta.  
The bandedge modes have discrete angular momenta that originate from different families of the parastichies whose numbers correspond to the Fibonacci numbers.  
The unique structural characteristic of the golden-angle spiral impose special features on the bandedge modes that are absent in the photonic crystals and quasicrystals.  
These modes may lead to unusual properties of light transport in the spiral structure, and also produce laser emission with well-defined angular momenta when optical gain is added.  
\section*{Acknowledgments}
This work is supported partly by the NSF grants DMR-0808937.  
This work is partially supported by the Air Force program "Deterministic Aperiodic Structures for On-chip Nanophotonic and Nanoplasmonic Device Applications" under Award FA9550-10-1-0019, and by the NSF Career Award No. ECCS-0846651. 

\end{document}